\documentclass[showpacs,amsmath,amssymb,floatfix,prl,twocolumn]{revtex4}
\usepackage{graphicx,amsmath}

\begin{document}


\title {Contactless  photoconductivity measurements on (Si) nanowires}

\author{ A.D. Chepelianskii$^{(a)}$, F. Chiodi$^{(a)}$, M. Ferrier$^{(a)}$, S. Gu\'eron$^{(a)}$, E. Rouviere$^{(b)}$ and H. Bouchiat$^{(a)}$ }
\affiliation{$(a)$ LPS, Univ. Paris-Sud, CNRS, UMR 8502, F-91405, Orsay, France }
\affiliation{$(b)$ CEA, LITEN/DTNM/LCRE, 17, rue des Martyrs, Grenoble, France }

\begin{abstract}
Conducting nanowires possess remarkable physical properties unattainable in bulk materials.
However our understanding of their transport properties is limited by the difficulty of connecting them electrically.
In this Letter we investigate phototransport in both bulk silicon and silicon nanowires using a superconducting 
multimode resonator operating at frequencies between $0.3$ and $3$ GHz. 
We find that whereas the bulk Si response is mainly dissipative, the nanowires exhibit a large 
dielectric polarizability. 
This technique is contactless and 
can be applied to many other semiconducting nanowires and molecules. 
Our approach also allows to investigate the coupling of electron transport to surface acoustic waves in bulk Si 
and to electro-mechanical resonances in the nanowires. 
\end{abstract}

\maketitle

In recent years transport properties of conducting nanowires attracted a considerable interest.
Synthesis of carbon nanotubes and semiconducting nanowires opened the possibility 
of new mechanical, electronic and optical applications. For example carbon nanotubes 
allowed to fabricate single electron transistors operating at room temperatures \cite{Dekker2001},
and very high quality nano-electromechanical resonators in suspended nanotube samples \cite{VanDerZant2009}.
Numerous photonics applications were achieved with Si nanowires including microcavities and waveguides \cite{Koshida2009}.
Despite these successes, our understanding of transport properties of nanowires is limited 
by our ability to make good ohmic contacts at low temperatures. 
Thus for carbon nanotubes the nature of electronic transport at low temperatures is still 
unknown (possibilities include formation of a Luttinger liquid or dynamical coulomb blockade) \cite{Bachtold}. 
For more exotic nanowires like DNA, even qualitative information on whether the molecule 
is conducting or insulating is not reliable \cite{Endres}.

In this letter, we propose a generic experiment to probe photoconductivity without direct contacts. 
To this end, we couple nanowires to a multimode electromagnetic (EM) resonator. 
Light irradiation is used to excite carriers in the nanowires which interact with the EM field 
of the resonator and change the resonance parameters. We demonstrate this technique in practice, 
by measuring photo-transport in bulk silicon and Si nanowires. 

\begin{figure}
\begin{center}
\includegraphics[clip=true,width=8cm]{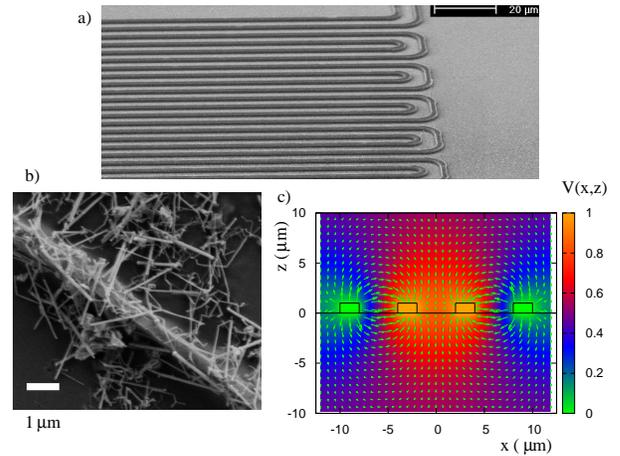}
\vglue -0.25cm
\caption{(Color online) a) Scanning electron microscope (SEM) image of the superconducting niobium meanders that form a multimode resonator. 
The substrate in this samples is Si/${\rm SiO_2}$.
b) SEM image of a resonator after deposition of undoped Si nanowires, for this experiment the substrate is sapphire.
c) FEM calculation of the electric field far from the meander boundaries.
The color/gray scale indicates the value of the scalar potential $V(x,z)$ while the arrows show the direction and amplitude 
of the electric field $\mathbf{E} = -{\mathbf \nabla} V$. 
}
\label{BlueRayFig1}
\end{center}
\end{figure}

Our probe is a multimode EM resonator formed by two superconducting meanders of total  
length $L_{R} \simeq 25\;{\rm cm}$ (see Fig.~\ref{BlueRayFig1}.a). This structure has regularly spaced resonant frequencies 
given by $f \simeq f_n = n f_1$ where $f_1 \simeq 365\;{\rm MHz}$ and $n$ is an integer. 
The meanders were fabricated by etching 
a $1\; \mu m$ thick niobium film with $SF_6$ reactive ion etching. 
During this procedure the meanders 
were protected by an aluminum mask patterned using optical lithography. 
In a last step $Al$ was dissolved in a KOH solution \cite{Reulet1,Reulet2}. 
Two types of dielectric substrates were used. 
In a first experiment the resonator was prepared on top 
of a Si/SiO$_2$ substrate (the oxide layer was $\simeq 500$nm thick),
which allowed us to probe conductivity of bulk Si under light irradiation (see Fig.~\ref{BlueRayFig1}.a).
In a second experiment we used sapphire as dielectric substrate 
and 
we deposited vapour-liquid-solid grown  Si nanowires \cite{ERouviere} on top of the resonator
(see Fig.~\ref{BlueRayFig1}.b). 
Note that sapphire remains insulating under blue irradiation (energy $2.5\;{\rm eV}$)
due to its high band gap of $9.9\;{\rm eV}$.

The coupling between the resonator and the Si-nanowires is greatly enhanced 
by the resonator's meander structure  which strongly confines the EM field 
near the substrate interface where the nanowires are deposited. 
We have checked this using a finite element (FEM) calculation of the potential 
and of the electric field. The results of the simulation are shown on Fig.~\ref{BlueRayFig1}.c, and 
clearly indicate that the electric field vanishes for $|z| \le 5\;{\rm \mu m}$ where 
$|z|$ is the distance to the substrate interface (distance between meanders is $D = 7\;{\rm \mu m}$). 
The FEM simulations also show that the electric field is mainly oriented along the interface for $z = 0$.  

\begin{figure}
\begin{center}
\includegraphics[clip=true,width=8.5cm]{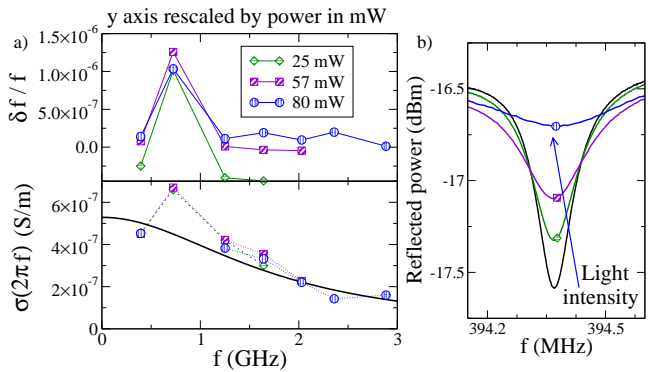}
\vglue -0.25cm
\caption{(Color online) Measurements on a bulk Si sample (see Fig.~\ref{BlueRayFig1}.a).
a) Top panel: relative shift of resonance frequencies under light irradiation as a function 
of microwave frequency, symbol abscissa correspond to resonance positions. Bottom panel: photo-conductivity $\sigma(2 \pi f)$ 
as obtained from the drop of resonance quality factor $Q$ using Eq.~(\ref{eq:SigmaOmega}).
The smooth black curve represents a Drude fit with a relaxation time of $\tau = 90\;{\rm ps}$. 
In both panels symbol shape indicates electrical power provided to the diode. 
The values of $\delta f/f$ and $\sigma$ are divided by the power value in mW and collapse on a single curve 
(at highest power data must be rescaled by 190mW to coincide with other curves, we attribute this to nonlinear dependence of 
light intensity on electrical power).
b) Broadening of the fundamental resonance under light irradiation (drop of $Q$ factor). Temperature is $4.2\;{\rm K}$. 
}
\label{BlueRayFig2}
\end{center}
\end{figure}

During the  measurements the resonators are immersed in liquid $He^{4}$ at $4.2\;{\rm K}$.
The shape and position of the resonances are determined 
by measuring the reflection along a coaxial cable capacitively coupled 
to the resonator. At resonant frequencies more power is absorbed by the resonator,
and a dip in reflected power is observed (incident microwave power was $-60\;{\rm dBm}$). 
The resonators were then irradiated with a blue light provided by a commercial diode operating at low temperature.
For samples on bulk Si the resonances are strongly broadened under irradiation,
a typical behavior is shown in Fig.~\ref{BlueRayFig2}.b for the first harmonic of the resonator 
while no broadening was observed for a control resonator on a sapphire substrate.

This drop of the resonance quality factor under irradiation can be understood in terms of photoinduced conductivity in Si. 
Indeed the photon energy of $\hbar \omega \simeq 2.5\;{\rm eV}$ is much larger than the Si gap $\Delta \simeq 1.2\;{\rm eV}$.
Thus the absorption of photons creates a stationary population of electron/holes pairs, and
a finite photo-conductivity $\sigma(2 \pi f)$ where $f$ is the microwave frequency. 
In the regime of a weak conductor with $\sigma \ll \epsilon\; 2 \pi f$ the screening is negligible and 
the associated drop in the resonator quality factor is well described by the relation \cite{Landau}: 
\begin{align}
{\rm Re}\; \sigma(2 \pi f) =  \epsilon \; 2 \pi f \; \delta Q^{-1} = \frac{\sigma_0}{ 1 + (2 \pi f \tau)^2}
\label{eq:SigmaOmega} 
\end{align}
Here $\epsilon$ is  dielectric constant in Si, and the last equality in Eq.~(\ref{eq:SigmaOmega}) 
is a Drude  approximation to $\sigma(2 \pi f)$ with elastic relaxation time $\tau$. 
The quantity $\delta Q^{-1}$ is the difference between $Q^{-1}$ with and without irradiation.
The experimental values for $\epsilon\; 2 \pi f\; \delta Q^{-1}$ are shown on Fig.~\ref{BlueRayFig2}, 
the quality factors are determined using both 
amplitude and phase of the reflection coefficient to remove 
the artificial broadening of the resonance induced by the coupling to the transmission line. 
Since the excited carrier concentration is proportional to light intensity,
we rescale the data of Fig.~\ref{BlueRayFig2} by the electrical power absorbed by the diode.

A good agreement with Eq.~(\ref{eq:SigmaOmega}) is found for a relaxation time of $\tau = 90\;{\rm ps}$
that corresponds to a mobility $\mu \simeq e \tau / m \simeq 10^{5}\;{\rm cm^2/(V s)}$ 
(here $m$ is the electron mass). From this value 
the effective carrier temperature $T_{eff} \simeq 20\;{\rm K}$ can be determined using 
the mobility dependence on temperature in high purity Si \cite{Canali}.
This effective temperature is determined by the energy transferred by a photon $\hbar \omega - \Delta$ to an electron-hole pair 
and energy dissipation mechanisms in Si. It does not depend on light power as long as the carrier density $n_+$ remains 
small (from conductivity data on Fig.~\ref{BlueRayFig2} we estimate $n_+ \simeq 10^8\;{\rm cm^{-3}}$). 
The Drude approximation works for all frequencies except for the second resonance at $f = f_2 \simeq 725\;{\rm MHz}$. 
The relative frequency shift under irradiation $\delta f/f$ has also a peak at $f \simeq f_2$.
The origin of this resonant frequency can be understood in terms of emission of surface acoustic waves (SAW)
in the Si substrate. Indeed the periodic structure of the meanders also creates a resonance for SAW.
The corresponding frequency is determined by the distance between meanders  $D \simeq 7\;{\mu m}$
and the transverse sound velocity in Si: $V_T \simeq 5500\;{\rm m/s}$ \cite{Wortman}, giving 
a resonant frequency of $f_{SAW} = V_T / D \simeq 785\;{\rm MHz}$ close to the frequency $f_2$ of the EM resonator. 
The blue irradiation intensity is also modulated with a period $D$ by the presence of the resonator and 
the continuous excitation of electron-hole pairs can create a charge density wave with period $D$ 
strongly increasing the coupling between the acoustic and EM modes. This would explain why SAW resonance appears very clearly 
in the photoconduction data. 

\begin{figure}
\begin{center} 
\includegraphics[clip=true,width=9cm]{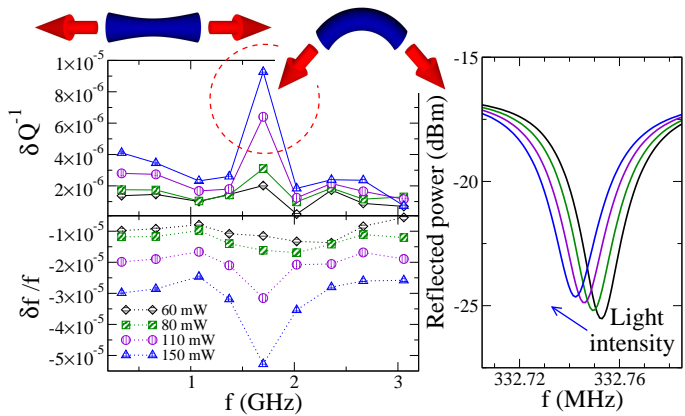}
\vglue -0.25cm
\caption{ (Color online) Measurements on the sample with Si nanowires (see also Fig.~\ref{BlueRayFig1}.b):
a) $\delta Q^{-1}$ and $\delta f/ f$ as a function of frequency 
for several light intensities, 
b) Shift and broadening of the fundamental resonance under light irradiation.
On control sample without Si nanowires typical $|\delta f/ f| \le 2\times10^{-6}$ and $|\delta Q^{-1}| \le 10^{-6}$ 
at maximum light power. 
}
\label{BlueRayFig3}
\end{center}
\end{figure}

We also used this technique to probe phototransport in Si nanowires (see sample on Fig.~\ref{BlueRayFig1}.b).
The typical behavior under irradiation is shown for the first resonance on Fig.~\ref{BlueRayFig3}. 
Contrarily to samples on bulk Si where only a broadening of the resonance was observed, 
a shift of the resonance position is clearly visible. 
In the bulk the charges were not confined in the direction parallel to the Si interface 
hence only dissipative photoconductive response $\propto \delta Q^{-1}$ could be observed.
In the nanowires, photo-induced charges are confined and can be polarized, 
creating a photopolarizability $\propto \delta f / f$.
%
%
In order to compare the magnitude of these two effects
we assume that we excite at most one electron-hole pair in a nanowire, and we call $n_{Si^+}$ the surface density of the excited nanowires. 
The polarization $P$ of an excited nanowire is determined from an equilibrium Boltzmann distribution 
in presence of the resonator electric field $E$: $P \simeq \int^{L/2}_{-L/2} q x \exp\left( \frac{q E x}{T_{eff}} \right) \frac{dx}{L}  \simeq q^2 L^2 E / T_{eff}$.
Here $L$ is the nanowire length, $q$ is the carrier charge and $T_{eff}$ is the effective carrier temperature.
This polarization creates a shift of the resonant frequencies:
\begin{align}
\frac{\delta f}{f} \simeq -\frac{q^2 L^2}{\epsilon T_{eff}} \frac{n_{Si^+}}{D} . 
\label{eq:DeltaF}
\end{align}
where the factor $1/D$ originates from the confinement of the EM field.
Photoconductivity is determined by Eq.~(\ref{eq:SigmaOmega}), the unknown conductivity $\sigma_0$ can be estimated with a Drude approximation:
$\sigma_0 \simeq \frac{q^2 \tau_{Si} n_{Si^+}}{m D}$ where $n_{Si^+}/D$ is the effective carrier concentration. The relaxation time $\tau_{Si}$ 
is determined by the collisions with the nanowire walls $\tau_{Si} \simeq R \sqrt{ m / T_{eff} }$ where $R$ is the nanowire radius and $m$ 
is the electron mass. Combining these results with Eq.~(\ref{eq:DeltaF}) yields the dimensionless ratio $\Gamma$:
\begin{align}
\Gamma = \delta Q^{-1} / (\delta f/ f) \simeq -\frac{R }{2 \pi f L^2} \sqrt{ \frac{T_{eff}}{m} }
\label{eq:Ratio}
\end{align}
The quantity $\Gamma$ does not depend on $n_{Si^+}$ and thus on light power. This is confirmed by extracting $\Gamma$ from the data of Fig.~\ref{BlueRayFig3}
which gives typical $\Gamma \simeq 10^{-1}$. After injecting this value in Eq.~(\ref{eq:Ratio}) together 
with $R \simeq 50\;{\rm nm}$, $L \simeq 2.5\;{\rm \mu m}$ and $f \simeq 400\;{\rm MHz}$ we find an effective temperature 
$T_{eff} \simeq 65\;{\rm K}$ of the same order of magnitude as in bulk Si.

The frequency dependence in Fig.~\ref{BlueRayFig3} is characterized by a peak at frequency $f_R \simeq 1.7\;{\rm GHz}$.  
The origin of this peak can be related to mechanical resonances in the nanowires, similarly to interaction with SAW for bulk Si.
Possible excited modes are shown on top of Fig.~\ref{BlueRayFig3}, and correspond to bending and stretching of the nanowire. For the bending mode 
resonance occurs when a transverse phonon wavelength fits in a nanowire. For an average nanowire length $L \simeq 2.5{\rm \mu m}$ 
(length fluctuations are around $1\;{\rm \mu m}$) 
the resonant frequency is $f_R \simeq V_T / L \simeq 2.2 \;{\rm GHz}$.
For the stretching mode, only half a wavelength enters the nanowire at resonance, hence the frequency is 
$f_R \simeq V_L / (2 L) \simeq 1.8\;{\rm GHz}$,
where $V_L \simeq 9000\;{\rm m/s}$ is the longitudinal sound velocity in Si \cite{Wortman}.  
The stretching mode gives a better frequency estimate which suggests that mainly this mode is excited in our experiment. 

In conclusion we have shown that using a high Q multimode resonator we can probe effectively low temperature photo-transport in bulk Si and Si nanowires. 
For bulk Si, photo-induced carriers induce a dissipative response which broadens the resonances.
The drop of quality factor allows to deduce the relaxation time and the carrier effective temperature.
At a special frequency resonant interaction with surface acoustic waves is observed.
In Si nanowires, photo-induced carriers can polarize the nanowires and thereby induce a dominant non dissipative response absent in the bulk.
We showed that the ratio between dissipative and non-dissipative responses determines the effective carrier temperature in the nanowires.
For some special resonant frequencies mechanical resonances in the nanowire could be excited.
We stress that this technique is very generic and can be applied to many other systems where photoconductivity is expected, possible examples 
include DNA and photochromic molecular switches. Coupled with optical spectroscopy, it could provide valuable transport data 
on nanowires with embedded quantum dots. 

We acknowledge ANR QuantADN and NanoTERA for support and thank S. de Franceschi, B. Semin and S. Saranga for help in experiment preparation.

\end{document}